# ANALYSIS OF POINT CONTACT SPECTRA OF $KFe_2As_2$ IN THE THERMAL REGIME

**N.V. Gamayunova, O.E. Kvitnitskaya, N.L.Bobrov, and Yu. G. Naidyuk**

*B. Verkin Institute for Low Temperature Physics and Engineering, National Academy of Sciences of Ukraine,*
*47 Lenin Ave., 61103, Kharkiv, Ukraine*

ABSTRACT

The point-contact spectra of the iron-based compound $KFe_2As_2$ were analyzed according to the thermal regime theory. We have obtained the values of the residual resistivity, the Lorentz number, and the electron mean free path in the contacts. It was shown that the most point-contact spectra can be described by this theory. The reasons for this behavior are discussed.

PACS: 73.40.-c, 74.70.Dd

Keywords: iron-based superconductors, point-contact spectroscopy, thermal regime

Recently the research of iron-based superconductors (IBS) has been one of the main directions in superconducting materials science [1]. Non-doped $KFe_2As_2$ takes a special place among IBS compounds as it is the final composition in the family of $Ba_{1-x}K_xFe_2As_2$ solid solutions of 122-group of IBS.

The system $Ba_{1-x}K_xFe_2As_2$ can attain high-temperature superconductivity ($T_C^{max}$=38K at x=0.4) due to chemical substitutions, when Ba-ions are chemically replaced for K-ions. Thus, $KFe_2As_2$, as well as $BaFe_2As_2$, can be considered the "parent" compounds for this family. Moreover, $BaFe_2As_2$ is not a superconducting material and exhibits the long-range antiferromagnetic order, while $KFe_2As_2$ is a superconductor with a low critical temperature ($T_C$=3.8K).

The knowledge of electronic, magnetic and structural properties of the «parent» 122-compounds is crucial for understanding the mechanism of superconductivity, which develops on doping these materials. Numerous investigations of these compounds have been intended to promote the synthesis of advanced materials with assigned properties for practical application. It has formed one of the priority areas in solid state physics [1].

It is of particular interest to investigate $KFe_2As_2$ by the method of point-contact spectroscopy (PCS) because point contacts (PC) make it possible to obtain the spectra of elementary excitations in conductors [2]. PC is a small-size (from a few nanometers to several micrometers) electrical contact, produced by two bulk electrodes forming a small-area contact.

In ballistic or diffusive PCs the spectroscopic regime of current flow is realized. In this case the electron mean free path is much longer than the size of a contact. The second derivative of the current-voltage characteristics (CVC) of these contacts contains information about the function of the electron-phonon interaction, as well as other collective boson interactions in the normal state [2].

Besides, a thermal regime of current flow [3, 4] can be realized, especially in PCs based on the compounds with high specific resistivity. In this case the inelastic electron mean free path is smaller than a diameter of the contact. As a result, the temperature in the contact increases with the applied voltage [3, 4]. The heating of the contact area determines the CVC nonlinearity.

Here we investigate the CVC derivatives of PCs formed on the basis of single crystal $KFe_2As_2$. First of all, it is of great interest to clear up the regime of current flow through the contacts.

The single crystals of $KFe_2As_2$ (lateral dimensions 1x0.5 $mm^2$, thickness ~ 0.1 mm) were grown using the flux - method in IFW (Dresden) [5]. The temperature dependence of the specific electrical resistivity has a typical metallic behavior with low residual resistivity (see Fig.1). The onset of the superconducting transition is slightly below 4 K (Fig.1, right-hand inset). The $KFe_2As_2$−Cu and $KFe_2As_2$−Ag PCs were formed using the conventional "needle-anvil" technique [2] by touching a cleaved single crystal surface with a sharpened thin Cu or Ag wire. The precision two-coordinate mechanism enables changing the pressing force of the electrodes and scanning the sample surface. The PCs were formed in a liquid helium cryostat at T=4.2 K, which is higher than the superconducting transition temperature in $KFe_2As_2$ ($T_C$ = 3.8 K).

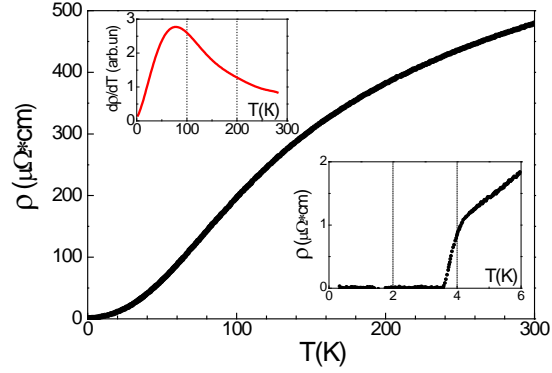

Fig. 1. The temperature dependences of electrical resistivity $\rho_{ab}(T)$ in the (100) plane of the single crystal $KFe_2As_2$. Left-hand inset: shape of the temperature derivative of the resistivity in the main panel. Right-hand inset: the superconducting transition as seen in the resistivity data.

The differential resistance $dV/dI(V)$ for such PCs and the second derivative $d^2V/dI^2(V)$ of the CVC were obtained by the standard detecting technique. The first and second harmonics of the modulating signal proportional to the corresponding derivatives of the CVC were measured with a lock-in amplifier.

The PC spectra $d^2V/dI^2(V)$ measured on the $KFe_2As_2$ sample can be divided into two groups. The first smaller group contains spectra with a pronounced maximum at 20 mV. It was suggested [6], that in this case the electrons pass through the contact in ballistic or diffusive regimes. The spectra of the other much larger group display a broad maximum from 30 to 60 mV and half-width of $40 \div 60$ mV (Fig. 2). The shapes of the $d^2V/dI^2$-curves are similar to the temperature derivative of the resistivity $d\rho/dT$ (see Fig.1, left-hand inset). The growth of the differential resistance $\Delta R/R$ amounts to 300%. It is also one of characteristics of the thermal regime, therefore we assume the thermal regime of current flow in this case.

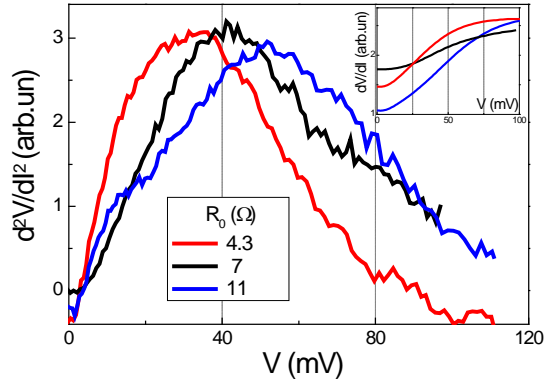

Fig. 2. PC spectra measured at T=4.2K for several thermal PCs based on $KFe_2As_2$. The inset shows the differential resistance for each curve from the main panel.

To confirm this assumption, we calculated CVCs for PCs based on $KFe_2As_2$ and their first derivatives $dV/dI$ within the theory of the thermal regime [3,4] and then compared the theoretical and experimental curves.

According to the theory of thermal regime [3,4], the temperature inside the contact $T_{PC}$ depends on the voltage applied to the contact as:

$$T_{PC}^2 = T_0^2 + \frac{V^2}{4L_0}, \qquad (1)$$

where $T_0$ is the bath temperature, $L_0 = 2.45 \cdot 10^{-8}$ V$^2$/K$^2$ is the Lorentz number.

In this case, the CVC of the PC is expressed by the formula [3,4]:

$$I(V) = Vd \int_0^1 \frac{dx}{\rho(T_{PC}(1-x^2)^{1/2})}. \qquad (2)$$

The CVC for our PCs and its first derivative dV/dI were obtained according to Eq.(2). The calculations were based on the temperature dependences of electrical resistivity $\rho_{ab}(T)$ in the (100) plane (Fig.1) and $\rho_c(T)$ along the axis *c* [7] of the KFe$_2$As$_2$ compound. To match the shapes and magnitudes of experimental and theoretical curves, the calculation was made by varying the following parameters: the Lorentz number L, the residual resistivity $\rho_0$ and the contact size d. It is seen in Fig. 3 that the theoretical dV/dI curve agrees well with the experimental one. The parameters used to simulate the dV/dI spectra are shown in Table.1 for several contacts.

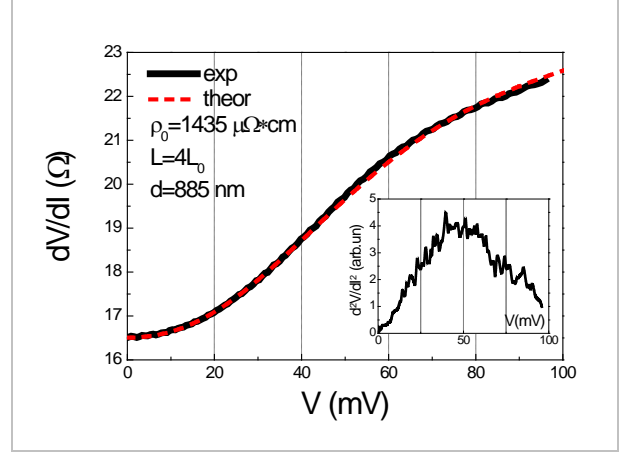

Fig. 3. The comparison of the theoretical (dashed red line) and experimental (full black line) dV/dI curves for the PC based on KFe$_2$As$_2$ (№1 in the Table 1).The parameters used to simulate the dV/dI-spectra are shown in the panel. The inset shows the d$^2$V/dI$^2$ spectrum for this contact.

The parameters L, $\rho_0$ and d were calculated for eight typical "thermal" spectra for PCs based on KFe$_2$As$_2$. They were: L≈4L$_0$, $\rho_0$ from several hundreds to thousands μΩ·cm, d about a few microns for PC resistance in the range of 2 to 10 Ω (Table 1).

Table 1. The parameters used to simulate the PC dV/dI-spectra for thermal PCs based on KFe$_2$As$_2$: the PC resistance R$_0$ at zero bias, the related growth of the PC resistance R/R$_0$, the reduced Lorenz number L/L$_0$, the residual resistivity $\rho_0$ of the KFe$_2$As$_2$ sample, the PC diameter d, and the electron mean free path *l*.

| № | R$_0$, Ω | R/R$_0$ | $\rho_0$, μΩ·cm | L/L$_0$ | d, nm | *l*, nm |
|---|---|---|---|---|---|---|
| 1 | 16,5 | 1,4 | 1435 | 4 | 885 | 0,3 |
| 2 | 7 | 1,4 | 1340 | 4 | 1940 | 0,3 |
| 3 | 3 | 1,5 | 900 | 3,3 | 3000 | 0,5 |
| 4 | 11,2 | 1,6 | 840 | 4 | 740 | 0,5 |
| 5 | 11 | 2,5 | 450 | 4 | 380 | 1,0 |
| 6 | 8,5 | 2,9 | 350 | 4 | 420 | 1,3 |
| 7 | 6 | 3,5 | 275 | 4 | 460 | 1,6 |
| 8 | 3,5 | 4 | 220 | 4 | 650 | 2,0 |

For investigated PCs we can estimate the electron mean free path *l* using the value $\rho l$ and residual resistivity $\rho_0$ in the contact. Within the free electron model we have the relation $\rho l \approx 1.3 \cdot 10^4 n^{-2/3}$ [2], where $n \approx 4.9 \cdot 10^{21}$ cm$^{-3}$ in KFe$_2$As$_2$. The value *n* was taken from the calculation of the number of electrons per unit cell [5]. The electron mean free path was 0.3nm÷2nm, which is comparable with the parameters of the tetragonal lattice of KFe$_2$As$_2$: $a \approx 0.4$ nm and $c \approx 1.4$ nm [5].

The question thus arises: why are the $\rho_0$ and d values so high and why is *l* so small? One can assume that the electron density of the KFe$_2$As$_2$ surface is smaller than that in the bulk. The reason may be that the stoichiometric composition of the compound is disturbed by the chemically reactive and volatile alkali metal K, i.e. K causes degradation of the sample surface.

If the electron density decreases to the values typical for the semimetals or heavily doped semiconductors (i.e. $n$ is $10^3$ times lower than $n \approx 4.9 \cdot 10^{21}$ cm$^{-3}$), the parameter $\rho l$ will increase by two orders of magnitude and $l$ will increase correspondingly. On the other hand, if $\rho l$ increases, the contribution of the Sharvin resistance will increase and the contribution of the Maxwell resistance will decrease according to the Wexler formula for the PC resistance.

$$R_W = \frac{\rho}{d} + \frac{16\rho l}{\pi d^2} \qquad (3)$$

The Wexler resistance is the sum of the Maxwell resistance $R_M$ (the first term) and the Sharvin resistance $R_{Sh}$ (the second term). In turn, such increase in $\rho l$ will result in $\rho_0$ increase.

The low value of the electron mean free path in the crystal may result from the disturbance of the crystalline structure of the sample surface in the process of the formation of the clamping contacts by the needle-anvil method at low temperatures. It is also may be the degradation of the surface.

The calculated curves of some contacts deviate from the experimental ones as is shown in Fig. 4. We therefore assume that the crystallite orientation of the contact area deviates from the basic plane and approaches the axis $c$. In this case the fit can be improved if our calculation allows for the contribution of the dependence $\rho_c(T)$ of KFe$_2$As$_2$ compound.

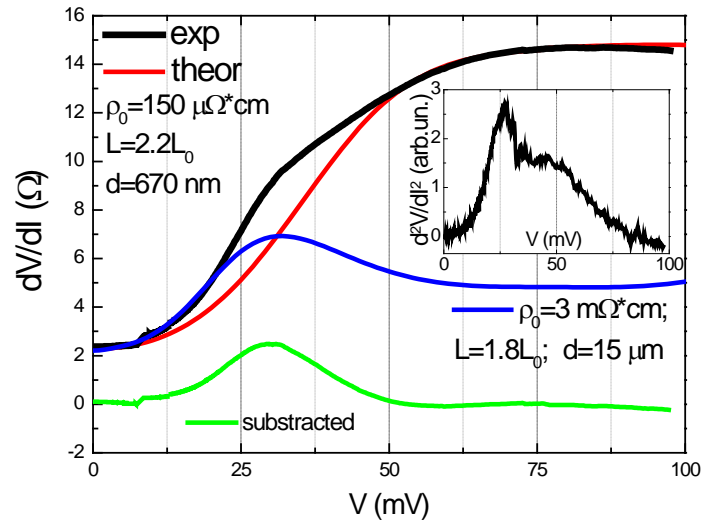

Fig. 4. The comparison of the experimental dV/dI-curve (black line) for the KFe$_2$As$_2$-Ag PC and the theoretical one (red line) based on the $\rho_{ab}(T)$ dependence. The green curve shows the difference between them. The blue curve is a calculated one basing on $\rho_c(T)$. It simulates the green curve. The parameters, used to simulate these dV/dI spectra, are shown in the panel. The inset shows the PC spectrum for this contact.

Thus, the results obtained indicate that the nonspectral thermal regime is mainly realized in the PCs based on KFe$_2$As$_2$. This must be taken into account in PCS measurements aimed at separating the thermal effects and obtaining the spectral information about quasi-particle excitations for the investigated and related isostructural compounds. Regarding the determination of the regime of the current flow in such contacts, there is still the question of building a physical model, which would be based not only on imperfect contact and /or degradation of the surface.

The authors thank S. Aswartham and S. Wurmehl for providing of KFe$_2$As$_2$ single crystals.

[1] A.A. Kordyuk, *Low Temp. Phys* **38**, 888 (2012).
[2] Yu.G. Naidyuk, I.K.Yanson, Point-Contact Spectroscopy. – NY: Springer, V.**145** (2005).
[3] B.I.Verkin, I.K.Yanson, I.O.Kulik, O.I.Shklyarevski, A.A.Lysykh, Yu.G.Naydyuk, *Solid State Commun*. **30,** 215 (1979).
[4] I.O. Kulik, *Fiz. Nizk. Temp*. **18**, 450 (1992) [*Sov. J. Low Temp. Phys* **18**, 302 (1992)].
[5] M.Abdel-Hafiez, S.Aswartham, S.Wurmehl, V.Grinenko, C.Hess, S.-L.Drechsler, S. Johnston, A.U.B.Wolter, and B.Büchner, H.Rosner, and L.Boeri, *Phys. Rev. B* **85**, 134533 (2012).
[6] Yu. G. Naidyuk, O. E. Kvitnitskaya, and N. V. Gamayunova, L. Boeri, S. Aswartham, S. Wurmehl, B. Büchner, D.V. Efremov, G. Fuchs, and S.-L. Drechsler, *Phys. Rev. B* **90**, 094505 (2014).
[7] T.Terashima, M.Kimata, H.Satsukawa, A.Harada, K.Hazama, S.Uji, H.Harima, G.-F.Chen, J.-L.Luo, and N.- L.Wang, *J. of the Physical Society of Japan* **78**, 063702 (2009).